\documentclass[prl,aps,twocolumn,psfig,preprintnumbers,amsmath,amssymb,showpacs]{revtex4}
\usepackage{graphicx}\usepackage{epsfig}
\usepackage{color}

\begin{document}

\title{Break-up fragment topology in statistical multifragmentation models}

\author{Ad. R. Raduta$^{1,2}$}
\affiliation{$^{1}$~NIPNE, Bucharest-Magurele, POB-MG6, Romania\\
$^{2}$~Institut de Physique Nucleaire, Universite Paris-Sud 11, CNRS/IN2P3, 
F-91406 Orsay cedex, France
}

\begin{abstract}
Break-up fragmentation patterns together with kinetic and configurational energy fluctuations
are investigated in the framework of a microcanonical model
with fragment degrees of freedom over a broad excitation energy range. 
As far as fragment partitioning is approximately preserved,
energy fluctuations are found to be rather insensitive to both 
the way in which the freeze-out volume is constrained and the trajectory
followed by the system in the 
excitation energy - freeze-out volume space.
Due to hard-core repulsion, the freeze-out volume is found to be populated un-uniformly,
its highly depleted core giving the source a bubble-like structure. 
The most probable localization of the largest fragments in the freeze-out volume 
may be inferred experimentally from their kinematic properties,
largely dictated by Coulomb repulsion. 

\end{abstract}


\maketitle

\section{I. Introduction}

For more than two decades, nuclear multifragmentation benefits from a constant 
scientific interest whose main motivation is the observation of a 
(liquid-gas-like) phase transition at sub-atomic scale \cite{springerbook,bb}.  

Relying on the presumptive existence of an equilibrated break-up stage in the
simultaneous multi-particle decay of excited nuclei, statistical models
with cluster degrees of freedom \cite{mmmc,smm,randrup,mmm,dasgupta}
represent particularly useful tools for the characterization of the
equilibrated state of the source and, not less important, the study of the
associated thermodynamics. The remarkable advantage of realistically
incorporating most properties of bound and continuum states via empirical
parameterizations of cluster energies or level densities explains their
ability to well describe a wealth of experimental data produced over a broad
energetic domain.

It was demonstrated that experimental data corresponding to a well-defined 
equilibrated source may be described by a
unique solution of such a statistical model \cite{mmm-prc2002,mmm-prc2005}.
It is nevertheless not true that the different statistical models converge to
the same equilibrated source if the analysis is done by exclusively considering
experimental (after-burner) information \cite{tsang-epja2006}.
This is partly due to the different thermodynamical constraints imposed to the
employed statistical ensembles or mathematical tricks designed in order to simplify 
the partition function or speed-up the simulation and, to a much larger
extend, to the differences in the break-up fragment definition.

The aim of the present work is to contribute to a deeper understanding of the
break-up stage of the multifragmentation decay as ruled by statistical laws.
For this reason, contributions from dynamics (as radial collective flow) and
sequential particle evaporations from primary fragments will be referred to only
tangentially, despite that over an important region of the considered energy domain
they play an important role. 
For the same reason we will ignore also eventual fragments recombination
subsequent to the break-up, thoroughly considered by some authors 
\cite{pal1995,de2005}.
More precisely, we want to see
\begin{itemize}
\item whether fluctuations of different energetic degrees of freedom are mainly
  dictated by the localization of the decay event into the phase diagram
  or, conversely, by the dominant fragmentation modes,
\item whether break-up nuclear matter distribution is uniform and, if not,
\item whether is it possible to trace the un-homogeneities from experimentally 
  accessible data.
\end{itemize}

The paper is organized as follows: Sec. II offers a brief review on the
statistical models of multifragmentation with a special focus on the
microcanonical ones, employed here;
Sec. III investigates the sharing of system's available energy among
different degrees of freedom and the sensitivity of the energy fluctuations to
the system phase properties and fragment partition; 
Sec. IV focuses on break-up patterns and the
extend in which these may be inferred from kinetic energy distributions. 
Modifications of fragment charge distributions 
brought by considering that,
at variance with the standard break-up picture, primary fragments interact through
nuclear forces are also addressed in Section IV.
Conclusions are drawn in Sec. V.

\section{II. Statistical treatment of multifragmentation}

Under the equilibrium hypothesis, statistical models reduce the physical
problem under study to the estimation of the number of microscopic states
compatible with the thermodynamical macroscopic constraints.
This implies that assuming that it is possible to write down the 
mathematical expression of the statistical weight of a configuration $W_C$
in the appropriate statistical ensemble, all the thermodynamic quantities may
be calculated out of the characteristic partition sum,
\begin{equation}
{\cal Z}=\sum_C W_C,
\end{equation}
while any ensemble-averaged observable $X$ may be expressed as,
\begin{equation}
<X_C>=\frac{\sum_C W_C X_C}{ \sum_C W_C}.
\end{equation}

While for relatively large extensive systems, thermodynamical properties are
not sensitive to the way in which the statistical ensemble is defined, 
when dealing with small systems, as the nuclear ones, it is important to
choose the most appropriate replica of the physical phenomenon.
The lack of any thermal or chemical potential reservoirs in the case of isolated 
multifragmenting nuclei, recommend the microcanonical ensemble 
as the most reasonable choice \cite{mmmc,gross2000,gross2001}. 
In this case, it is obvious that the conserved quantities are
the total protons ($Z$) and neutrons ($A-Z$) numbers, 
the total energy ($E$), total momentum (${\rm \bf P}$) 
and, eventually, total angular momentum (${\rm \bf L}$).
The freeze-out volume ($V$) may be considered either as fixed, either as
fluctuating. 

Defining a generic break-up configuration by the isotopic, internal and
translational properties of each fragment,
$C=\{ A_1, Z_1, \epsilon_1, {\rm\bf r}_1, ..., A_{N_C}, Z_{N_C},
\epsilon_{N_C}, {\rm\bf r}_{N_C}\}$,
one gets for the statistical weight of the constant volume ensemble 
the equation \cite{mmm},
\begin{eqnarray}
\nonumber
W_C(A,Z,E,V) \propto \frac1{N_C!} \Omega \prod_{n=1}^{N_C}\left( 
\frac{\rho_n(\epsilon_n)}{h^3}(mA_n)^{3/2}\right)
\\ 
~ \frac{2\pi}{\Gamma(3/2(N_C-2))} ~ \frac{1}{\sqrt{({\rm det} I})}
~ \frac{(2 \pi K)^{3/2N_C-4}}{(mA)^{3/2}},
\label{eq:wc}
\end{eqnarray}
where $I$ is the moment of inertia,
$K$ is the thermal kinetic energy and
 $\Omega=\chi V^{N_C}$ stays for the free volume or, equivalently, accounts for
inter-fragment interaction in the hard-core idealization. 
From Eq. (\ref{eq:wc}) it is straightforward to calculate the statistical weight
of a microcanonical ensemble with fluctuating volume as,
\begin{equation}
{\cal W}_c (A,Z,E,\lambda) =\int  W_C(A,Z,E,V) \exp(-\lambda V) dV.
\end{equation}

It is worthwhile to mention at this point that working under a
fixed total energy constraint,
it results that the thermal kinetic energy, a key thermodynamic quantity
related to the temperature through 
$T^{-1}=\left( \partial S/\partial E \right)=
1/W(A,Z,E,V) \partial
W(A,Z,E,V)/\partial E $,
is determined by the amount of the energy available after extracting from the
source excitation the costs
of fragment formation $\sum_i B_i$, fragment internal excitation 
$\sum_i \epsilon_i$
and mutual fragment 
interaction ($\sum_{i<j} V_{ij}$),
\begin{equation}
K=E_{ex}-Q-\sum_i \epsilon_i-\sum_{i<j} V_{ij}.
\label{eq:Kth}
\end{equation}
This implies that also the fluctuations of $K$ are strongly dependent on the
fluctuations of the other three energetic degrees of freedom, as we shall see
later on. 

The results discussed hereafter have been obtained in the framework of the
Microcanonical Model of Multifragmentation (MMM) \cite{mmm} in the case of the
medium size nucleus (130,60) within
the commonly accepted scenario according to which
the break-up fragments do not interact otherwise than via Coulomb forces.
The consequences of considering in the spirit of
Refs. \cite{pal1995,de2005,satpathy90} that break-up fragments feel also
the nuclear proximity potential are discussed only with respect to fragment
charge distributions, for the sake of completeness.
Despite the particular choices regarding the model and the nucleus, 
the results are considered generic for the statistical
break-up of multifragmenting nuclei.

\begin{figure}
\begin{center}
\includegraphics[angle=0, width=0.99\columnwidth]{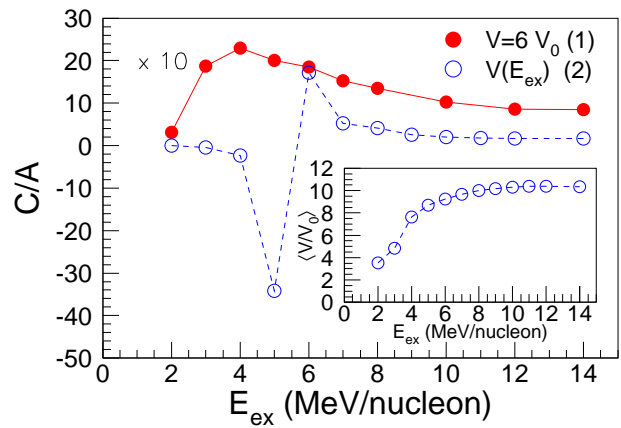}
\end{center}
\caption{(Color online)
Heat capacity versus source excitation energy for the multifragmenting nucleus
(130,60) which evolves through the phase space following
(1) a constant freeze-out volume $V=6V_0$ path, or, 
(2) a path characterized by an average freeze-out volume which increases with
the excitation as indicated in the inset.}
\label{fig:cv}
\end{figure}

Two arbitrary paths in the phase diagram have been considered, 
a constant volume path $V=6V_0$ (full symbols)
and one along which the {\em average} volume increases with excitation energy (open symbols). 
The motivation of choosing a constant volume path is twofold.
On one hand, it reproduces the fixed freeze-out volume statistical constrained
which was used for 
treating multifragmentation over almost two decades and, on the other hand, 
it accounts for the belief that the freeze-out volume (average) value does not change 
significantly while increasing source excitation energy.
Twofold is also the motivation of choosing the second path.
First, it cancels the statistical constraint of constant volume and, secondly,
it accounts for a freeze-out volume whose (average) value may increase with energy, 
as recent analyses of experimental data indicate \cite{silvia,eric}.
In this last case, the average freeze-out volume increases from 3.5$V_0$ at 2
MeV/nucleon to about 10.4$V_0$ at 14 MeV/nucleon, as indicated in the inset of
Fig. \ref{fig:cv}.
Even more importantly, the two paths differ by the regions of the system phase
diagram they explore. 
Thus, following the evolution of the heat capacity,
\begin{eqnarray}
  C^{-1}&=&-T^2 \left( \frac{\partial^2S}{\partial E^2}\right) \nonumber \\
  &=&1-T^2 \frac{1}{W(A,Z,E,V)} \frac{\partial^2 W(A,Z,E,V)}{\partial E^2}
  \nonumber \\
  &=&1-T^2\left<
    \frac{\left(\frac32 N-4\right)
      \left(\frac32 N-5\right)}{K^2}
  \right>,
\label{eq:C}
\end{eqnarray}
plotted in Fig. \ref{fig:cv} as a function of excitation energy, one may notice
that the constant volume path is supra-critical, while the 
increasing average-volume path crosses the phase coexistence region. 
Phase coexistence is signaled by negative values of the heat capacity.

\section{III. Energy sharing at break-up: average and RMS values}

\begin{figure}
\begin{center}
\includegraphics[angle=0, width=1.1\columnwidth]{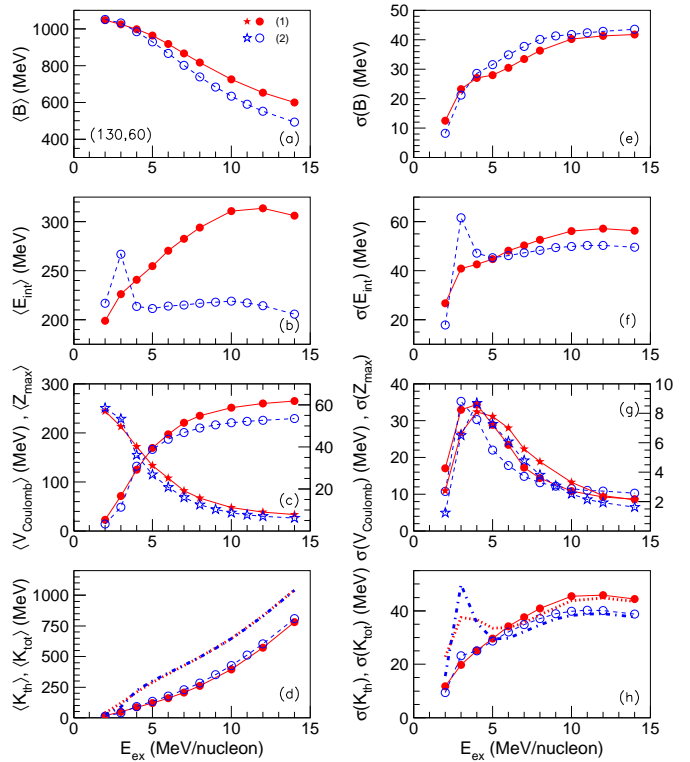}
\end{center}
\caption{(Color online)
Evolution with source excitation energy of mean (left column) and RMS (right column)
values of total binding energy (upper panel), total fragment internal
excitation (second upper panel),
Coulomb fragment-fragment interaction (third panel) and 
thermal kinetic energy (bottom panel) corresponding to the break-up stage 
of the (130, 60) multifragmenting nucleus. 
The considered states along a constant volume, $V=6 V_0$, (1) and 
an average volume increasing with excitation (2) paths 
are represented with solid and, respectively, open circles.
Third raw panels depict also, with open and solid stars, the excitation energy
dependence of the largest fragment in each decay event.
The dashed lines in the bottom panel stand for total fragment kinetic energy.}
\label{fig:fluct}
\end{figure}

Fig. \ref{fig:fluct} presents the average values of total fragment binding
energy (upper panel), 
internal excitation (second upper panel),
Coulomb interaction (third upper panel) and thermal kinetic energy 
(bottom panel) (left column) 
together with their RMS values (right column)
corresponding to the break-up stage of a (130,60) nucleus whose excitation
energy ranges from 2 to 14 MeV/nucleon along the two considered trajectories.
Distributions of the mean charge of the largest fragment and its RMS are 
superimposed on the third upper panels with full and open stars.
Dashed lines are used in the bottom panels to indicate how total fragment kinetic energy, 
a quantity experimentally accessible, behaves with respect to source excitation.

One can see that, irrespectively the considered path, the more and more
advanced fragmentation allowed by an
increasing source energy, suggested by a rapidly decreasing $Z_{max}$,
leads to a monotonic diminish of the total binding energy 
and a monotonic increase of the total Coulomb interaction energy.
The total binding energy decrease is due to the increasing
fragment surfaces while the increase of the total Coulomb interaction energy
is explained by an increasingly uniform occupation of the volume. 
While the curves corresponding to the two considered paths diverge with excitation, 
they are still not far
one from another, as for a given $E_{ex}$
their values differ by at most 20\% in the considered energy domain.
In contrast with this, the amount of energy dissipated in fragment internal
excitation has a more complex
evolution and the relative difference among the values obtained along the two paths reaches 
50\% at $E_{ex}$=14 MeV/nucleon.
Nevertheless, the evolution and relative magnitude of the above quantities are such that the 
kinetic energy increases monotonically, as one would expect (see left bottom panel).

The right panels of Fig. \ref{fig:fluct}
present the energy fluctuations and indicate that, as more and more fragment 
partitions 
are possible with the increasing energy, $\sigma(B)$ and $\sigma(E_{int})$
rise as well. 
Very interestingly, $\sigma(V_C)$ augments up to 4 MeV/nucleon and then decreases. 
The positive slope region corresponds to the energy domain where configurations containing 
one heavy residue are dominant.
The negative slope interval corresponds to a regime of rather advanced fragmentation
which allows for a more uniform population of the freeze-out volume.
As one may notice, the peak of $\sigma(V_C)$ corresponds roughly to the peak
of $\sigma(Z_{max})$ 
(full and open stars in the third right panel)
and indicates that the largest fragment $Z_{max}$
dictates the geometrical arrangement of fragments and, finally, the Coulomb energy.

Another remark is that, because of the fact that the reduction of 
Coulomb energy fluctuation is less significant 
than the increase of internal excitation and binding energy fluctuations, 
$\sigma(K_{th})$ increases monotonically.
Nevertheless, analysing the experimentally accessible fragment total kinetic
energy distribution, one would note a peak at 3 MeV/nucleon, 
as the consequence of summing up the peaked  $\sigma(V_C)$ with the
monotonically increasing  $\sigma(K_{th})$ (right bottom panel).

But, the first important result is that fluctuations of kinetic and configurational
energetic channels prove rather insensitive to both freeze-out volume
constraints and
the trajectory followed by the system into the excitation energy - freeze-out volume plane,
provided that the fragmentation pattern is preserved.
The result is striking the more as the two considered trajectories explore different
regions of the phase diagram.

\section{IV. Fragmentation patterns and nuclear matter radial distributions}

\begin{figure}
\begin{center}
\includegraphics[angle=0, width=0.95\columnwidth]{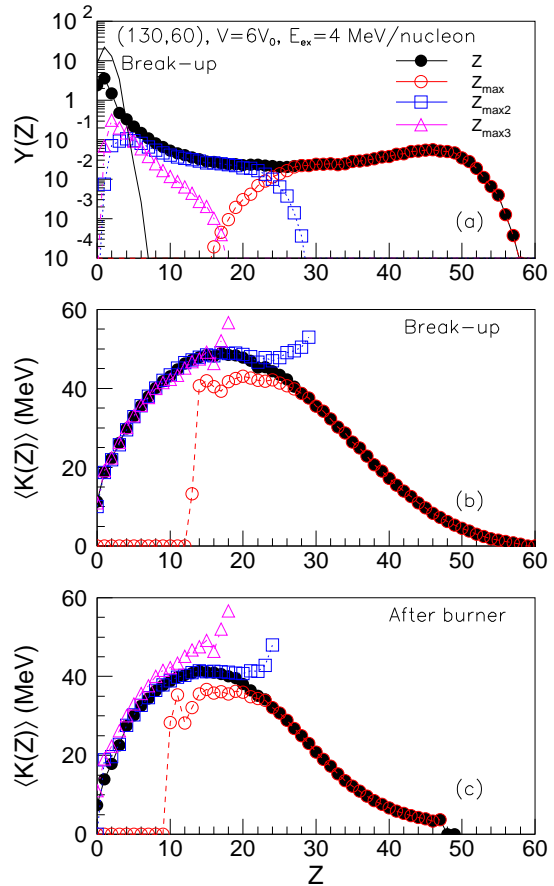}
\end{center}
\caption{(Color online) Charge distributions (top) and 
distributions of average kinetic energy as a function of fragment charge
(middle, bottom) corresponding to the multifragmentation
of the (130,60) nucleus with $V=6 V_0$ and $E_{ex}$= 4 MeV/nucleon.
The two upper panels depict break-up stage results, while after-burner data
are illustrated in the bottom panel.
Solid circles stand for all fragments while open circles,
open squares and open triangles stand for the
largest, second largest and, respectively, third largest fragment in each
event.
The solid line on the top panel corresponds to the
break-up fragment charge distribution obtained under the assumption that
break-up fragments interact not only through repulsive hard-core and Coulomb
potentials, but also via proximity potentials.
}
\label{fig:kz_ex=4}
\end{figure}

\begin{figure}
\begin{center}
\includegraphics[angle=0, width=0.95\columnwidth]{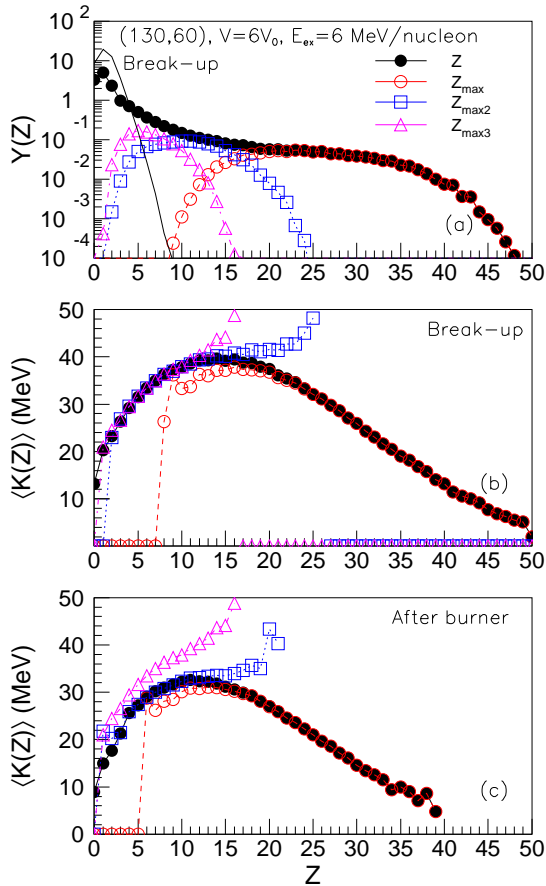}
\end{center}
\caption{(Color online)
The same as in Fig. \ref{fig:kz_ex=4} for the multifragmentation
of the (130,60) nucleus with $V=6 V_0$ and $E_{ex}$= 6 MeV/nucleon.}
\label{fig:kz_ex=6}
\end{figure}

The break-up fragmentation pattern corresponding to 4 MeV/nucleon excitation
energy, where   
the largest fluctuations in $\sigma(Z_{max})$ and $\sigma(V_C)$ manifest themselves,
is illustrated in the upper panel of Fig. \ref{fig:kz_ex=4} 
while the upper panel of Fig. \ref{fig:kz_ex=6} presents the 
fragmentation pattern obtained at a slightly
higher source excitation, 6 MeV/nucleon. 
As no sensitivity was found to the way in which the freeze-out volume in
constrained, from here on we shall consider only the case corresponding to
 $V=6 V_0$.

One can see that at $E_{ex}=4$ MeV/nucleon the dominant fragmentation mode
is characterized by a residue representing 80\% of total system but 
multifragmentation configurations are already possible.
For instance, 
configurations characterized by two intermediate size fragments ($Z_{max} \approx 30$ and
$Z_{max2} \approx 20$), though five times less probable than the most probable
fragmentation mode,
are nevertheless frequent enough to induce a quite flat $Y(Z_{max})$. 
The diversity of fragmentation modes translated in broad $Z_{max}$ and
$Z_{max2}$ 
distributions persists at 6 MeV/nucleon, but there is no more possible to identify
a close competition among different fragmentation patterns. 
This means that there are no more distinct manners of filling up the available volume, 
whose co-existence
leads to large fluctuation of the Coulomb energy.

We remind at this point that fragmentation patterns are nevertheless very
sensitive to break-up fragments definition or 
modelisation of the break-up stage itself.
If, for instance, one sticks to the non-interacting break-up fragments scenario
but considers that, in agreement with Thomas-Fermi
calculations, excited nuclei at freeze-out are diluted, the fragment charge
distribution will be settled by the competition between the reduced free
volume and the augmented thermal kinetic energy.
The same qualitative situation is reached if, not the fragments density but, 
their internal excitation is modified. 
If, for example, one adopts for the nuclear level density an expression which
leads to lower fragment internal excitation, in view of Eq. (\ref{eq:Kth}) 
$K_{th}$ will increase, favoring an increased reaction products multiplicity.
This last quantity, in its turn, by making possible a more uniform population of the
freeze-out volume characterized by a larger $V_C$, will tend to diminish $K_{th}$.

Much dramatic modifications are expected if one considers that break-up
fragments interact not only via repulsive Coulomb, 
but also via attractive nuclear proximity potentials. 
This conceptually different approach is mainly justified
by the fact that for break-up volumes of the order of few $V_0$
the distances between fragment surfaces may be inferior than $\sim$ 1 fm. 
This situation has been discussed recurrently by Das, De, Samaddar, Satpathy, Bonasera
and collab. \cite{pal1995,de2005,satpathy90} together with break-up fragment subsequent
recombination and shown to lead to an increased productivity of light and
heavy fragments at the cost of the intermediate ones.
As recombination, which occurs if two fragments approach
each other during the Coulomb propagation,
acts in the sense of washing-out the statistical properties of break-up
fragment formation,
here we shall restrict ourselves to comment exclusively on the consequences 
of modifying fragments energetics.

It is relatively easy to anticipate from Eq. (\ref{eq:Kth}) that, by
considering an extra attractive potential, one will get an increase of thermal kinetic energy
and reaction products multiplicity.
The confirmation is given by the solid curves on the top panels of 
Figs. \ref{fig:kz_ex=4} and \ref{fig:kz_ex=6} obtained
in the case in which the nuclear interactions are implemented as in
Ref. \cite{pal1995}.
In both situations one may notice a dramatic enhancement of the
light cluster multiplicity and the total supression of fragments 
with $Z \geq 10$.
These steep $Y(Z)$ distributions and the evolution of their slopes 
with source excitation may be reconciled with experimental data
if and only if one assumes
that final fragment formation is
dominated by post-break-up dynamics (including collective flow) 
and multiparticle correlation \cite{de2005}.
If this were the case, the freeze-out would occur much later
that the break-up.
The complete modelisation of this process is nevertheless a challenging task 
which goes beyond the goal of the present paper.

In addition to charge distributions, fragment average kinetic energy
distributions represent robust 
and directly accessible experimental information and make up 
a key ingredient in the standard procedure of identifying the statistically equilibrated 
source by confrontation with predictions of statistical models \cite{mmm-prc2002,mmm-prc2005}. 
The middle and bottom panels of Figs. \ref{fig:kz_ex=4} and \ref{fig:kz_ex=6} depict 
the average kinetic energy distributions of primary and, respectively, cold
fragments corresponding to the same source (130,60) with $V=6V_0$ and 
$E_{ex}=$ 4 and 6 MeV/nucleon.
The inclusive distributions are plotted with full circles, while
distributions corresponding to the largest, second largest and third largest fragment
are plotted with open circles, squares and triangles. 
Collective radial flow is set to zero to keep fragments statistical properties
unaffected.

Fragment average kinetic energy distributions are qualitatively similar for the two
source excitations. 
As one may notice, the maximum value of 49 (42) MeV reached by the primary (asymptotic) 
$<K(Z)>|_{4~{\rm MeV/nucleon}}$ distribution
exceeds by 25\% (30 \%) the maximum value obtained by the corresponding 
$<K(Z)>|_{6~{\rm MeV/nucleon}}$ distribution.
This result, in apparent contrast with what one would expect given the increase
with 57\% of the total Coulomb energy over the considered energy domain, may be understood
taking into account the much stronger increase in the total number of reaction products
\cite{mmm-plb2005}.

A common and interesting feature is present in the charge domain where the
$<K(Z)>$ distributions reach their maximum.
Thus, the break-up and asymptotic average kinetic energies 
of the largest fragment are systematically smaller than the 
average kinetic energies
of the second largest fragment which are, in their turn, smaller than the ones
corresponding to the third largest fragment. 
This result has been already pointed out by the Indra collaboration in the case of
Xe+Sn at 32 MeV/nucleon and Gd+U at 36 MeV/nucleon reactions \cite{tabacaru}
and shown to diminish
with the source excitation energy, in perfect agreement with the present results.
Taking into account that fragment kinetic energies are to a large extent dictated by Coulomb,
it becomes obvious that analyzing them one may get information on the most
probable fragment position at break-up.
Having the same dependence as Coulomb on fragment mass and distance from the
freeze-out volume center, 
radial collective flow, if present, would enhance this shift. 
If this reasoning is correct, it means that larger a fragment is, 
closer to the freeze-out volume center it is produced. 

\begin{figure}
\begin{center}
\includegraphics[angle=0, width=0.99\columnwidth]{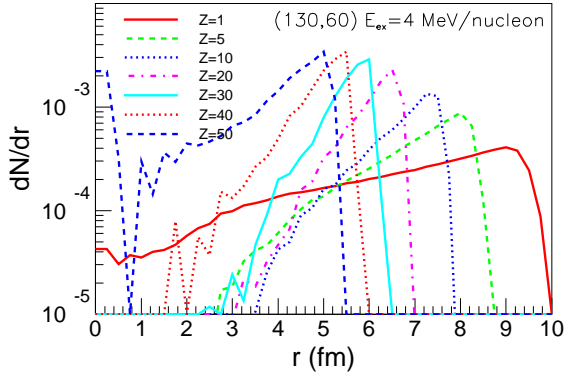}
\end{center}
\caption{(Color online)
Radial probability distributions of different size ($Z$=1, 5, 10, 20, 30, 40 and 50) 
primary fragments at break-up. 
The statistically equilibrated source (130, 60) is characterized by an excitation 
energy of 4 MeV/nucleon and a freeze-out volume $V=6V_0$.}
\label{fig:radial_fragment}
\end{figure}

The answer to this issue is offered by Fig. \ref{fig:radial_fragment}
 where radial probability distributions of 
different size fragments corresponding to the break-up stage of the 
(130,60) source with $V=6V_0$ and $E_{ex}=4$ MeV/nucleon are plotted. 
As a first general remark, one may say that the probability to create a fragment inside the 
freeze-out volume, whatever its size, is highly un-uniform and strongly
diminishes in the core region.
Moreover, heavy fragments are localized preferentially towards the inner parts, while relatively
light nuclei may be created over wider regions. 
This means that lighter is a fragment, stronger will be the Coulomb repulsion the charged
core will exercise over it and, consequently, higher its final kinetic energy.
This explains the observed systematic shift between the maximum values of kinetic energy 
corresponding to the three largest fragments.
The systematic reduction of the volume accessible to a fragment as its mass
increases is the consequence of the employed non-overlapping condition between a fragment 
and the wall of the container which mimics
the freeze-out volume. 
In the case of the heaviest fragments ($Z$=50), this geometric condition
is responsible for fragment concentration in a region which represents
only 15\% of the total freeze-out volume.
We remind that the classification of multifragmentation events with respect to fragments' 
spatial arrangement and its influence on fragment-fragment correlation functions
was discussed for the first time in Ref. \cite{schapiro} in the framework of
MMMC model \cite{mmmc}, where the authors have identified 'sun'- and 'soup'-like events.

\begin{figure}
\begin{center}
\includegraphics[angle=0, width=0.99\columnwidth]{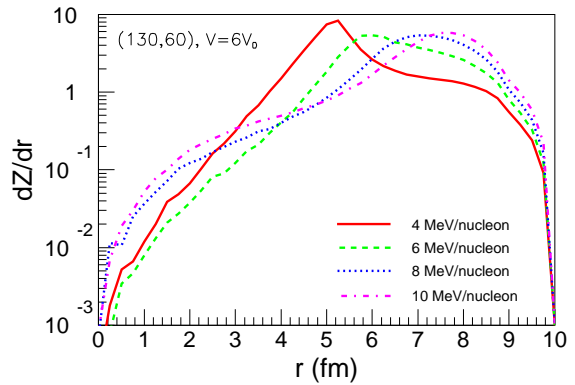}
\end{center}
\caption{(Color online)
Total charge radial distributions corresponding to the (130,60) multifragmentating nucleus
with $V=6V_0$ and different excitation energies (4, 6, 8 and 10 MeV/nucleon).}
\label{fig:radial_totalcharge}
\end{figure}

As the source excitation energy increases and fragmentation becomes more advanced, 
a more uniform population of the freeze-out volume is expected, 
such that the largest fragments kinetic energy shifts become negligible.
The evolution of the total charge radial distribution with source excitation is illustrated in 
Fig. \ref{fig:radial_totalcharge} for the same multifragmenting nucleus, (130, 60) with 
$V=6 V_0$. 
Indeed, at 8 MeV/nucleon the matter in the inner regions of the freeze-out volume 
is 10 times denser that the one produced at 6 MeV/nucleon, but the overall distribution remains
strongly outwards peaked, giving the source a bubble-like structure.
Bubble-like structures of the nuclear matter at break-up have been obtained
also in the framework of
stochastic mean-field approaches \cite{smf} which explain fragmentation on behalf
of growing volume and surface instabilities encountered during the expansion phase of the
excited system, as recently reported in Refs. \cite{colonna,parlog}.
This agreement between results of statistical models with cluster degrees of freedom and
dynamical models with nucleonic degrees of freedom is far from being trivial taking into account
the conceptually different scenarios the two categories of models advance for explaining
multifragmentation and the almost complementary treatment of the physical process.

\section{V. Conclusions}

To conclude, using a microcanonical multifragmentation model with cluster degrees of freedom
we have analyzed the break-up fragmentation patterns of a medium size equilibrated source
who follows different paths through the excitation energy - freeze-out volume space.
The constraints imposed on the freeze-out volume 
are found to not affect significantly the magnitude of different energy fluctuations.
Moreover, kinetic and configurational energy fluctuations are insensitive
to the system phase properties
as far as the considered fragment partitions are similar.
Over the whole domain of excitation energy, spatial matter distribution
at break-up is highly un-uniform, its outward peaked shape giving the source a 
bubble-like structure.
The most probable localization of nuclear fragments at break-up depends on
fragment mass and, because
of Coulomb acceleration, it is possible to infer it from the experimentally accessible 
fragment average kinetic energy distributions, especially at intermediate
values of source excitation. 
Thus, heavy fragments are found to be produced in the inner
regions of the freeze-out volume, while the lighter ones are produced in a larger region 
of the freeze-out volume.
Considering that break-up fragments interact not only through 
repulsive hard-core and Coulomb potentials but also
via proximity potentials, one obtains dramatic modifications of the break-up
fragmentation patterns which suggest that final fragment formation is strongly
influenced by post-break-up dynamics and multiparticle correlations.

\section{acknowledgements}
The author acknowledges partial support from the Romanian National Authority for Scientific
Research under PNCDI2 programme, grant {\it IDEI nr. 267/2007}.


\begin{references}

\bibitem{springerbook} "Dynamics and Thermodynamics with Nuclear Degrees
                        of Freedom", Eur. Phys. J. {\bf A30} (2006).
\bibitem{bb} B. Borderie and M. F. Rivet, Progr. Part. Nucl. Phys. {\bf 61}, 551 (2008).
\bibitem{mmmc} D. H. E. Gross, Rep. Progr. Phys. {\bf 53}, 605 (1990).
\bibitem{smm} J. P. Bondorf, A. S. Botvina, A. S. Iljinov, I. N. Mishustin and K. Sneppen,
              Phys. Rep. {\bf 257}, 133 (1995).
\bibitem{randrup} S. E. Koonin and J. Randrup, Nucl. Phys. A {\bf 474}, 173 (1987).
\bibitem{mmm} Al. H. Raduta and Ad. R. Raduta, Phys. Rev. C {\bf 55}, 1344
  (1997).
\bibitem{dasgupta} S. Das Gupta and A. Z. Mekjian,  Phys. Rev. C {\bf 57}, 1361
  (1998); S. Pratt and S. Das Gupta,  Phys. Rev. C {\bf 62}, 044603
  (2000).
\bibitem{mmm-prc2002} Al. H. Raduta and Ad. R. Raduta, Phys. Rev. C 65, 054610 (2002).
\bibitem{mmm-prc2005} Ad. R. Raduta, E. Bonnet, B. Borderie, N. Le Neindre, and M. F. Rivet,
                      Phys. Rev. C 72, 057603 (2005).
\bibitem{tsang-epja2006} M. B. Tsang {\it et al.}, Eur. Phys. J. {\bf A30},
  129 (2006).
\bibitem{pal1995} S. Pal, S. K. Samaddar, A. Das and J. N. De,
  Nucl. Phys. {\bf A586}, 466 (1995).
\bibitem{de2005} S. K. Samaddar, J. N. De and A. Bonasera, Phys. Rev. C 
{\bf 71}, 011601(R) (2005). 
\bibitem{gross2000} D.H.E. Gross and E. Votyakov, Eur. Phys. J, {\bf B15}, 115 (2000).
\bibitem{gross2001} D.H.E. Gross, Nucl. Phys. {\bf A681}, 366 (2001).
\bibitem{satpathy90} L. Satpathy, M. Mishra, A. Das and M. Satpathy,
  Phys. Lett. {\bf B237}, 181 (1990).
\bibitem{silvia} S. Piantelli {\it et al.}, Nucl. Phys. {\bf A809}, 111 (2008).
\bibitem{eric} E. Bonnet and the INDRA collab., Nucl. Phys. {\bf A816}, 1 (2009).
\bibitem{mmm-plb2005} Ad. R. Raduta, B. Borderie, E. Bonnet, N. Le Neindre,
  S. Piantelli, M. F. Rivet,
Phys. Lett. {\bf B623}, 43 (2005). 
\bibitem{tabacaru} G. Tabacaru {\it et al.}, Nucl. Phys. {\bf A764}, 371 (2006).
\bibitem{schapiro} O. Schapiro and D.H.E. Gross, Nucl. Phys. {\bf A576}, 428 (1994).
\bibitem{smf} A. Guarnera, M. Colonna, Pf. Chomaz, Phys. Lett. {\bf B373}, 297 (1996);
M. Colonna {\it et al.}, Nucl. Phys. {\bf A642}, 449 (1998).
\bibitem{parlog}   M. Parlog, G. Tabacaru, J. P. Wieleczko, J. D. Frankland,
  B. Borderie, A. Chbihi, 
M. Colonna and M. F. Rivet, Eur. Phys. J. {\bf A25}, 223 (2005).  
\bibitem{colonna} M. Colonna, G. Fabbri, M. Di Toro, F. Matera and H.H. Wolter,
                  Nucl. Phys. {\bf A724}, 337 (2004).

\end{references}
\end{document}